\documentclass{bmcart}
\usepackage{amsmath,graphicx,amsfonts,epsfig}
\usepackage{times}
\usepackage{amsbsy}
\usepackage{latexsym}
\usepackage{amssymb}

\def\includegraphics{}

\startlocaldefs
\endlocaldefs
\begin{document}

\begin{frontmatter}

\begin{fmbox}
\dochead{Research}

\title{Review of Buffer-Aided Distributed Space-Time Coding Schemes and Algorithms for Cooperative Wireless Systems}

\author[
   addressref={aff1},                   
   corref={aff1},                       
   email={jg849@york.ac.uk}   
]{\inits{J}\fnm{Jiaqi} \snm{Gu}}
\author[
   addressref={aff1,aff2},
   email={rodrigo.delamare@york.ac.uk}
]{\inits{R. C.}\fnm{Rodrigo C.} \snm{de Lamare}}

\address[id=aff1]{
  \orgname{Department of Electronics}, 
  \street{University of York},                     %
  \postcode{YO10 5DD}                                
  \city{York},                              
  \cny{UK}                                    
}
\address[id=aff2]{%
  \orgname{CETUC},
  \street{PUC-Rio},
  \cny{Brazil}
}

\end{fmbox}

\begin{abstractbox}
\begin{abstract}
In this work, we propose buffer-aided distributed space-time coding
(DSTC) schemes and relay selection algorithms for cooperative
direct-sequence code-division multiple access (DS-CDMA) systems. We
first devise a relay pair selection algorithm that can form relay
pairs and then select the optimum set of relays among both the
source-relay phase and the relay-destination phase according to the
signal-to-interference-plus-noise ratio (SINR) criterion. Multiple
relays equipped with dynamic buffers are then introduced in the
network, which allows the relays to store data received from the
sources and wait until the most appropriate time for transmission. {
A greedy relay pair selection algorithm is then developed to reduce
the high cost of the exhaustive search required when a large number
of relays are involved.} The proposed techniques effectively improve
the quality of the transmission with an acceptable delay as the
buffer size is adjustable. An analysis of the computational
complexity of the proposed algorithms, the delay and a study of the
greedy algorithm are then carried out. Simulation results show that
the proposed dynamic buffer-aided DSTC schemes and algorithms
outperform prior art.
\end{abstract}

\end{abstractbox}

\end{frontmatter}

\section*{Introduction}

The ever-increasing demand for performance and reliability in
wireless communications has encouraged the development of numerous
innovative techniques. {  Cooperative diversity is one of the key
techniques that has been considered in recent years
\cite{Proakis,Ikki,Wei,Ikki1} as an effective tool to improving
transmission performance and system reliability. Several cooperative
schemes have been proposed \cite{sendonaris,Venturino,laneman04},
and among the most effective ones are Amplify-and-Forward (AF),
Decode-and-Forward (DF) \cite{laneman04,Kalantari,Nasab} and various
distributed space-time coding (DSTC) techniques
\cite{Wornell,Yiu,RCDL3,Jing1,Peng,healy2016design}.} For an AF
protocol, relays cooperate and amplify the received signals with a
given transmit power. With the DF protocol, relays decode the
received signals and then forward the re-encoded message to the
destination. DSTC schemes exploit spatial and temporal transmit
diversity by using a set of distributed antennas. With DSTC multiple
redundant copies of data are sent to the receiver to improve the
quality and reliability of data transmission. Applying DSTC at the
relays provides multiple processed signal copies to compensate for
the fading, helping to achieve the attainable diversity and coding
gains so that the interference can be more effectively mitigated. As
a result, better performance can be achieved when appropriate signal
processing and relay selection strategies are adopted.

\subsection*{Prior and Related Work}

{  In cooperative relaying systems, strategies that employ multiple
relays have been recently introduced in
\cite{Jing,Clarke,Talwar,Jiaqi1,Peng11,Jiaqi2,peng2016adaptive}. The
aim of relay selection is to find the optimum relay so the signal
can be transmitted and received with increased reliability.
Recently, a new cooperative scheme with relays equipped with buffers
has been introduced and analyzed in
\cite{Zlatanov1,Zlatanov2,Krikidis,Ikhlef}. In \cite{Zlatanov1}, a
brief introduction to buffer-aided relaying protocols for wireless
network is described and some practical challenges are discussed. A
further study of the throughput and diversity gain of the
buffer-aided system has been subsequently introduced in
\cite{Zlatanov2}. In \cite{Krikidis}, a selection technique that is
able to achieve the full diversity gain by selecting the strongest
available link in every time slot is detailed. In \cite{Ikhlef}, a
max-max relay selection (MMRS) scheme for half-duplex relays with
buffers is proposed. In particular, relays with the optimum
source-relay links and relay-destination links are chosen and
controlled for transmission and reception, respectively.}

\subsection*{Contributions}

{  In this work, we propose buffer-aided DSTC schemes and relay pair
selection algorithms for cooperative direct-sequence code-division
multiple access (DS-CDMA) systems. In the proposed cooperative
schemes, a relay pair selection algorithm that selects the optimum
set of relays according to the signal-to-interference-plus noise
ratio (SINR) criterion is performed at the initial stage. In
particular, if an exhaustive search is performed then all relay
pairs are examined and compared, while for the proposed greedy relay
pair selection a reduced number of relay pairs are evaluated.
Therefore, a link combination associated with the optimum relay
group is then selected, which determines if the corresponding
buffers are ready for either transmission or reception. After that,
the data transmission of the cooperative DS-CDMA system begins. In
particular, the direct transmission from the first phase occurs
between the source and the selected relay combination when the
buffers are in the reception mode. On the other hand, when the
corresponding buffers are switched to the transmission mode, the
proposed DSTC scheme is performed for each user from the selected
relay combination to the destination during the second phase. With
dynamic buffers equipped at each of the relays, the proposed
buffer-aided schemes take advantage of the storage of multiple
blocks of data so the most appropriate ones can be selected at a
suitable time instant. The key advantage of introducing the dynamic
buffers in the system is their ability to store multiple blocks of
data according to a chosen criterion so that the most appropriate
ones can be selected at a suitable time instant with the highest
efficiency. Furthermore, when referring to the cooperative DS-CDMA
systems, the problem of multiple access interference (MAI) that
arises from nonorthogonal received waveforms in DS-CDMA systems
needs to be faced. The use of buffers and relay selection can
effectively help in the interference mitigation by allowing
transmissions performed over channels with better propagation
conditions.}

The contributions of this paper are summarized as follows:
\begin{itemize}
\item We propose a buffer-aided DSTC scheme that is able to store enough data packets
in the corresponding buffer entries according to different criteria so that more appropriate symbols can be selected in a suitable time instant.

\item { We propose a relay selection algorithm that chooses a relay pair
rather than a single relay as the DSTC scheme needs the cooperation
of a pair of antennas.}

\item A greedy relay pair selection technique is then introduced to reduce
the high cost brought by the exhaustive search that is required when a large
number of relays are involved in the transmission.

\item We propose a dynamic approach so that the buffer size is adjustable
according to different situations.

\item An analysis of the computational complexity, the average delay and the
greedy algorithm are also presented.
\end{itemize}

The rest of this paper is organized as follows. In Section II, the
system model is described. In Section III, the dynamic buffer-aided
cooperative DSTC schemes are explained. In Section IV, the greedy
relay pair selection strategy is proposed. In Section V, the dynamic
buffer design is given and explained. The computational complexity
is studied, the analysis of the delay and the greedy algorithm are
then developed in Section VI. In Section VII, simulation results are
presented and discussed. Finally, conclusions are drawn in Section
VIII.

\section*{DSTC Cooperative DS-CDMA System Model}


We consider the uplink of a synchronous DS-CDMA system with $K$
users, $L$ relays equipped with finite-size buffers capable of storing $J$ packets and $N$ chips per
symbol that experiences channels with flat fading. The system is equipped with a cooperative protocol at each relay and we assume that the
transmit data are organized in packets comprising $P$ symbols. The
received signals are filtered by a matched filter and sampled at chip
rate to obtain sufficient statistics.
As shown in Fig. 1, the whole transmission is divided into two phases. In
the first phase, the source transmits the data to each of the relay over two consecutive time instants,
the decoded data over two time slots, $\hat{b}_{r_ld,k}(2i-1)$ and $\hat{b}_{r_ld,k}(2i)$, is stored at relay $l$
and is prepared to send data to the destination. A DSTC scheme is then employed at the following phase,
where the corresponding $2 \times 2$ Alamouti \cite{Alamouti,Chee,Liu} detected symbol matrix over relay $m$ and relay $n$ for user $k$ among two consecutive time instants is given by
\begin{equation}\label{equation0}
\textbf{B}_k = \left[\begin{array}{c c}
\hat{b}_{r_md,k}(2i-1) & -\hat{b}^*_{r_nd,k}(2i) \\
\hat{b}_{r_nd,k}(2i) & \hat{b}^*_{r_md,k}(2i-1) \\
 \end{array}\right].
\end{equation}
Consequently, the received signal for user $k$ from relay $m$ and $n$ to the destination
over two consecutive time slots yields the $2N\times1$ received vectors
described by
\begin{equation}\label{equation1}
\textbf{y}_{r_{m,n}d}(2i-1) =  \textbf{h}_{r_md}^k
\hat{b}_{r_md,k}(2i-1)+ \textbf{h}_{r_nd}^k
\hat{b}_{r_nd,k}(2i)+\textbf{n}(2i-1),
\end{equation}
\begin{equation}\label{equation2}
\textbf{y}_{r_{m,n}d}(2i)= \textbf{h}_{r_nd}^k
\hat{b}^*_{r_md,k}(2i-1)- \textbf{h}_{r_md}^k
\hat{b}^*_{r_nd,k}(2i)+\textbf{n}(2i),
\end{equation}
where $\textbf{h}_{r_ld}^k=a_{r_ld}^k\textbf{s}_k h_{r_ld,k}$ denotes an $N \times 1$ effective signature vector for user $k$ from the $l$-th relay to the destination with $m,n
\in [1,2,...,L]$. The quantity $a_{r_ld}^k$ represents the $k$-th user's
amplitude from the $l$-th relay to the destination,
$\textbf{s}_k=[s_k(1),s_k(2),...s_k(N)]^T$ is the $N\times1$ signature sequence
for user $k$ and $h_{r_ld,k}$ are the complex channel fading coefficients for user $k$
from the $l$-th relay to the destination. The $N\times1$ noise vectors
$\textbf{n}(2i-1)$ and $\textbf{n}(2i)$ contain samples of zero mean complex
Gaussian noise with variance $\sigma^2$, $\hat{b}_{r_ld,k}(2i-1)$ and
$\hat{b}_{r_ld,k}(2i)$ are the decoded symbols at the output of relay $l$ after
using a cooperative protocol at time instants $(2i-1)$ and $(2i)$,
respectively. Equivalently, (\ref{equation1}) and (\ref{equation2}) can be
rewritten as
\begin{equation}
\textbf{y}_{r_{m,n}d}=\textbf{H}_{r_{m,n}d}^k \textbf{b}_{r_{m,n}d,k}+\textbf{n}_{r_{m,n}d},
\end{equation}
where $\textbf{y}_{r_{m,n}d}=\left[\textbf{y}^T_{r_{m,n}d}(2i-1),(\textbf{y}^*_{r_{m,n}d}(2i))^T
\right]^T$ represents the received signal from relay $m$ and $n$ over two time
instants. The 2 N x 2 Alamouti matrix with the effective signatures for user k is given by
\begin{equation}
\textbf{H}_{r_{m,n}d}^k=\left[\hspace{-0.5em} \begin{array}{l}  \
\textbf{h}_{r_md}^k \ \ \ \ \ \ \ \ \  \textbf{h}_{r_nd}^k \\
\\
(\textbf{h}_{r_nd}^k)^* \ \ \ -(\textbf{h}_{r_md}^k)^* \\
\end{array}\hspace{-0.5em} \right],
\end{equation}
the $2\times1$ vector $\textbf{b}_{r_{m,n}d,k}=\left[\hat{b}_{r_md,k}(2i-1),
\hat{b}_{r_nd,k}(2i)\right]^T$ is the processed vector when the DF protocol is
employed at relays $m$ and $n$ at the corresponding time instant, and
$\textbf{n}_{r_{m,n}d}=\left[  \textbf{n}(2i-1)^T, (\textbf{n}^*(2i))^T \right]^T$ is
the noise vector that contains samples of zero mean complex
Gaussian noise with variance $\sigma^2$.

{  At the destination, various MUD schemes can be employed. For
linear MUD, the detected symbols can be obtained as given by
\begin{equation}
\hat{\textbf{b}}_{r_{m,n}d,k}=Q \big( (\textbf{w}^k_{r_{m,n}d})^H \textbf{y}_{r_{m,n}d}  \big)
\end{equation}
where $\textbf{w}^k_{r_{m,n}d}$ is the receive filter for user $k$
at the destination.

Similarly, maximum likelihood (ML) detection can also be applied at
the destination. The ML detector solves the following problem
\begin{equation}
\tilde{\textbf{b}}_{r_{m,n}d,k} = {\rm arg \ min} \| \textbf{y}_{r_{m,n}d}- \textbf{H}_{r_{m,n}d}^k \textbf{b}_{r_{m,n}d,k} \|^2,
\end{equation}
and the symbols obtained by the ML algorithm when the Alamouti
scheme is used are computed as given by
\begin{equation}
\begin{split}
&\tilde{b}_{r_md,k}(2i-1)=(\textbf{h}_{r_md}^k)^H \textbf{y}_{r_{m,n}d,k}(2i-1)+(\textbf{h}_{r_nd}^k)^T \textbf{y}^*_{r_{m,n}d,k}(2i)\\
&=\Big( (\textbf{h}_{r_md}^k)^H \textbf{h}_{r_md}^k + (\textbf{h}_{r_nd}^k)^T (\textbf{h}_{r_nd}^k)^* \Big) \hat{b}_{r_md,k}(2i-1)+  \Big( (\textbf{h}_{r_md}^k)^H \textbf{n}(2i-1) + (\textbf{h}_{r_nd}^k)^T \textbf{n}^*(2i) \Big)\\
\\
&\tilde{b}_{r_nd,k}(2i)=(\textbf{h}_{r_nd}^k)^H \textbf{y}_{r_{m,n}d,k}(2i-1)- (\textbf{h}_{r_md}^k)^T \textbf{y}^*_{r_{m,n}d,k}(2i)\\
&=\Big( (\textbf{h}_{r_nd}^k)^H \textbf{h}_{r_nd}^k + (\textbf{h}_{r_md}^k)^T (\textbf{h}_{r_md}^k)^* \Big) \hat{b}_{r_nd,k}(2i) + \Big(
(\textbf{h}_{r_nd}^k)^H \textbf{n}(2i-1) - (\textbf{h}_{r_md}^k)^T \textbf{n}^*(2i) \Big)\\
\end{split}
\end{equation}
Consequently, after testing all possible symbols for ML detection,
the most likely detection symbols are selected.} This scheme groups
the relays into different pairs and a more reliable transmission can
be achieved if proper relay pair selection is performed.

\section*{Proposed Buffer-aided Cooperative DSTC Scheme}

In this section, we present a buffer-aided cooperative DSTC scheme,
where each relay is equipped with a buffer so that the processed
data can be stored and the buffer can wait until the channel pair
associated with the best performance is selected. Consequently,
processed data are stored at the corresponding buffer entries and
then re-encoded when the appropriate time interval comes.
Specifically, the buffer with size $J$ can store up to $J$ packets
of data and can either forward or wait for the best time instant to
send data. This method effectively improves the quality of the
transmission, guarantees that the most suitable signal is selected
from the buffer entries and sent to the destination with a higher
reliability.

The algorithm begins with a SINR calculation for all possible
channel combinations. In the case of the Alamouti code, every two
relays are combined into a group and all possible lists of
corresponding channel pairs are considered. { Thus, when referring
to the SINR for a specific relay pair, the signal transmitted or
received by other relays are seen as interference. } The
corresponding SINR is then calculated and recorded as follows:
\begin{equation}\label{equation3}
\scriptsize{{\rm SINR}_{sr_{m,n}}=\frac{\sum\limits_{k=1}^K \textbf{w}_{s_kr_m}^H \rho_{s_kr_m} \textbf{w}_{s_kr_m}+\textbf{w}_{s_kr_n}^H \rho_{s_kr_n} \textbf{w}_{s_kr_n} }{\sum\limits_{k=1}^K \sum\limits_{\substack{l=1\\l\neq m,n}}^L \textbf{w}_{s_kr_l}^H \rho_{s_kr_l} \textbf{w}_{s_kr_l} + \sigma^2\textbf{w}_{s_kr_m}^H \textbf{w}_{s_kr_m}+ \sigma^2\textbf{w}_{s_kr_n}^H \textbf{w}_{s_kr_n}},}\vspace{-0.25cm}
\end{equation}
\begin{equation}\label{equation4}
\scriptsize{{\rm SINR}_{r_{m,n}d}=\frac{\sum\limits_{k=1}^K (\textbf{w}_{r_md}^k)^H \rho_{r_md}^k \textbf{w}_{r_md}^k+(\textbf{w}_{r_nd}^k)^H \rho_{r_nd}^k \textbf{w}_{r_nd}^k}{\sum\limits_{k=1}^K \sum\limits_{\substack{l=1\\l\neq m,n}}^L (\textbf{w}_{r_ld}^k)^H \rho_{r_ld}^k \textbf{w}_{r_ld}^k + \sigma^2(\textbf{w}_{r_md}^k)^H \textbf{w}_{r_md}^k+ \sigma^2(\textbf{w}_{r_nd}^k)^H \textbf{w}_{r_nd}^k},}\vspace{-0.25cm}
\end{equation}
where $\rho_{s_kr_l}=\textbf{h}_{s_kr_l}^H \textbf{h}_{s_kr_l}$ is the correlation coefficient of the desired user $k$ between the source and relay $l$, $\rho_{r_ld}^k=(\textbf{h}_{r_ld}^k)^H \textbf{h}_{r_ld}^k$ is the correlation coefficient for user $k$ from relay $l$ to the destination. $\textbf{h}_{s_kr_l}=a_{s_kr_l}\textbf{s}_kh_{s_kr_l}$ is the channel vector from user $k$ to relay $l$. In Eq. (\ref{equation3}), ${\rm SINR}_{sr_{m,n}}$ denotes the SINR for the combined paths from all
users to relay $m$ and relay $n$, $\textbf{w}_{s_kr_l}$ is the detector used at the relays. When the RAKE receiver is adopted at the corresponding relay, $\textbf{w}_{s_kr_l}$ is expressed as
\begin{equation}
\textbf{w}_{s_kr_l}=\textbf{h}_{s_kr_l},
\end{equation}
similarly, if the linear minimum mean-square error (MMSE) receiver
\cite{RCDL5,int,Chen,Meng,l1cg,zhaocheng,alt,jiolms,jiols,jiomimo,jidf,fa10,saabf,barc,honig,mswfccm,song,locsme}.
is employed at the relays, $\textbf{w}_{s_kr_l}$ is equal to
\begin{equation}
\textbf{w}_{s_kr_l}=\bigg(\sum\limits_{k=1}^K \textbf{h}_{s_kr_l}\textbf{h}^H_{s_kr_l}+\sigma^2 \textbf{I}\bigg)^{-1}\textbf{h}_{s_kr_l},
\end{equation}
$\textbf{h}_{s_kr_l}=a_{s_kr_l}\textbf{s}_k h_{s_kr_l}$ is the effective signature vector
from user $k$ to the relay $l$. Similarly, in Eq. (\ref{equation4}), ${\rm SINR}_{r_{m,n}d}$ represents the
SINR for the combined paths from relay $m$ and relay $n$ to the destination. The receiver  filter $\textbf{w}_{r_ld}^k$ is employed by the detector used at the destination. When the RAKE receiver is adopted at the destination, $\textbf{w}_{r_ld}^k$ is expressed as
\begin{equation}
\textbf{w}_{r_ld}^k=\textbf{h}_{r_ld}^k.
\end{equation}
Similarly, if the linear MMSE receiver is employed at the relays, $\textbf{w}_{r_ld}^k$ is equal to
\begin{equation}
\textbf{w}_{r_ld}^k=\bigg(\sum\limits_{k=1}^K \textbf{h}_{r_ld}^k(\textbf{h}_{r_ld}^k)^{H}+\sigma^2 \textbf{I}\bigg)^{-1} \textbf{h}_{r_ld}^k.
\end{equation}
The above equations correspond to a cooperative system  under the
assumption that signals from all users are transmitted to the
selected relays $m$ and $n$. Both RAKE and MMSE receivers
\cite{jpais,sicdma,zhang2015large} are considered here for the
purpose of complexity, it should be mentioned that other detectors
\cite{RCDL4,mfsic,mbdf,did,uchoa2016iterative} and precoders
\cite{mbthp,} can also be used. We then sort all these SINR values
in a decreasing order and select the one with the highest SINR as
given by
\begin{equation}
{\rm SINR_{p,q}={\rm arg}\max _{\substack{m,n\in[1,2,...,L]}} \{ {\rm SINR}_{sr_{m,n}},{\rm SINR}_{r_{m,n}d} \}}, \label{equation5}
\end{equation}
where ${\rm SINR_{p,q}}$ denotes the highest SINR associated with the relay $p$ and relay $q$.
After the highest SINR corresponding to the combined paths is selected,
two different situations need to be considered as follows.\\
\\
\textbf{Source-relay link}:
\vspace{3mm}

If the highest SINR belongs to the source-relay link, then the signal sent to
the target relays $p$ and $q$ over two time instants is given by
\begin{equation}\label{equation6}
\textbf{y}_{sr_l}(2i-1)=\sum\limits_{k=1}^K \textbf{h}_{s_kr_l}b_k(2i-1)+\textbf{n}(2i-1), l\in[p,q],
\end{equation}
\begin{equation}\label{equation7}
\textbf{y}_{sr_l}(2i)=\sum\limits_{k=1}^K \textbf{h}_{s_kr_l}b_k(2i)+\textbf{n}(2i),l\in[p,q].
\end{equation}
The received signal is then processed by the detectors as the DF protocol is
adopted. Therefore, the decoded symbols that are stored and sent to the
destination from the $l$-th relay are obtained as
\begin{equation}
\hat{b}_{r_ld,k}(2i-1)=Q(\textbf{w}_{s_kr_l}^{H}\textbf{y}_{sr_l}(2i-1)),
\end{equation}
and
\begin{equation}
\hat{b}_{r_ld,k}(2i)=Q(\textbf{w}_{s_kr_l}^{H}\textbf{y}_{sr_l}(2i)),
\end{equation}
where $Q(\cdot)$ denotes the slicer. After that, the buffers
are switched to the reception mode, the decoded symbol is consequently
stored in the corresponding buffer entries. Clearly, these operations are
performed when the corresponding buffer entries are not full, otherwise, the
second highest SINR is chosen as given by
\begin{equation}\label{equation8}
{\rm SINR^{pre}_{p,q}}={\rm SINR_{p,q}}
\end{equation}
\begin{equation}\label{equation9}
{\rm SINR_{u,v}} \in {\rm max} \{ {\rm SINR_{sr_{m,n}}},  {\rm SINR_{r_{m,n}d}}\} \setminus {\rm SINR^{pre}_{p,q}},
\end{equation}
where ${\rm SINR_{u,v}}$ denotes the second highest SINR associated with the
updated relay pair $\Omega_{u,v}$. $\{ {\rm SINR_{sr_{m,n}}},  {\rm
SINR_{r_{m,n}d}}\} \setminus {\rm SINR^{pre}_{p,q}}$ denotes a complementary
set where we drop the ${\rm SINR^{pre}_{p,q}}$ from the link SINR set $\{ {\rm
SINR_{sr_{m,n}}},  {\rm SINR_{r_{m,n}d}}\}$. Consequently, the above process
repeats in the following time instants.
\\
\\
\textbf{Relay-destination link}:
\vspace{3mm}

If the highest SINR is selected from the relay-destination link, in the
following two consecutive time instants, the buffers are switched to
transmission mode and the decoded symbol for user $k$ is re-encoded with the
Alamouti matrix as in (\ref{equation0}) so that DSTC is performed from the
selected relays $p$ and $q$ to the destination as given by
\begin{equation}\label{equation10}
\textbf{y}_{r_{p,q}d,k}(2i-1)= \textbf{h}_{r_pd}^k \hat{b}_{r_pd,k}(2i-1)+
\textbf{h}_{r_qd}^k \hat{b}_{r_qd,k}(2i)+\textbf{n}(2i-1),
\end{equation}
\begin{equation}\label{equation11}
\textbf{y}_{r_{p,q}d,k}(2i)= \textbf{h}_{r_qd}^k \hat{b}^*_{r_pd,k}(2i-1)-
\textbf{h}_{r_pd}^k \hat{b}^*_{r_qd,k}(2i)+\textbf{n}(2i).
\end{equation}
The received signal is then processed by the detectors at the destination.
Clearly, the above operation is conducted under the condition that the
corresponding buffer entries are not empty, otherwise, the second highest SINR
is chosen according to (\ref{equation8}) and (\ref{equation9}) and the above
process is repeated.

It is worth noting that for the purpose of simplicity, the above technique employed fixed-size buffers at the relays
so that the transmission delay can be controlled with accurate estimation.
The key advantage of the proposed scheme is its ability to select the most
appropriate symbols before they are forwarded to the next phase. In practice,
the performance highly depends on the buffer size $J$, the number of users $K$
and the accuracy of the detection at the relays. The proposed buffer-aided cooperative DSTC scheme is detailed in Table 1.

\begin{table}[!htb]
\centering\caption{The proposed buffer-aided cooperative DSTC scheme}
\begin{tabular}{l}
\hline
\% List all possible relay pairs\\
\% Select the combination with the highest SINR\\
${\rm SINR_{p,q}=max \{ {\rm SINR}_{sr_{m,n}},{\rm SINR}_{r_{m,n}d} \}}$\\
\%Source-relay link\\ \ \ \ \
\textbf{if} ${\rm SINR_{p,q}} \in [{\rm SINR_{sr_{m,n}}}], m,n \in [1,L]$\\\ \ \ \ \ \ \ \
\textbf{if} the buffers entries are not full\\ \ \ \ \ \ \ \ \ \ \ \ \
$\textbf{y}_{sr_l}(2i-1)=\sum\limits_{k=1}^K \textbf{h}_{s_kr_l} b_k(2i-1)+\textbf{n}_{sr_l}(2i-1),l\in[p,q],$\\ \ \ \ \ \ \  \ \ \ \ \ \
$\textbf{y}_{sr_l}(2i)=\sum\limits_{k=1}^K \textbf{h}_{s_kr_l} b_k(2i)+\textbf{n}_{sr_l}(2i),l\in[p,q].$\\ \ \ \ \ \ \ \ \ \
\%Apply the detectors at relay $n$ and relay $q$ to obtain\\\ \ \ \ \ \ \ \ \ \ \ \ \
$\hat{b}_{r_ld,k}(2i-1)$ and $\hat{b}_{r_ld,k}(2i)$ and store them \\ \ \ \ \ \ \ \ \ \ \ \ \ \
 in the corresponding buffer entries ($l\in[p,q]$) \\ \ \ \ \ \ \ \ \ \ \ \ \ \
\textbf{break}\\ \ \ \ \ \ \
\textbf{ else} \%choose the second highest SINR\\ \ \ \ \ \ \ \ \ \ \ \ \ \ \
${\rm SINR^{pre}_{p,q}}={\rm SINR_{p,q}}$\\ \ \ \ \ \ \ \ \ \ \ \ \ \ \
${\rm SINR_{p,q}} \in {\rm max} \{{\rm SINR_{sr_{m,n}}}, {\rm SINR}_{\rm r_{m,n}d} \} \setminus {\rm SINR^{pre}_{p,q}}$\\ \ \ \ \ \ \ \ \
\textbf{end}\\ \ \ \
\textbf{else} \%Relay-destination link\\ \ \ \ \ \ \ \ \ \ \ \ \ \ \
 ${\rm SINR_{p,q}} \in [{\rm SINR_{r_{m,n}d}}], m,n \in [1,L]$\\\ \ \ \ \ \ \ \ \
\textbf{if} the buffers entries are not empty\\ \ \ \ \ \ \ \ \ \ \ \ \ \
$\textbf{y}_{r_{p,q}d,k}(2i-1)= \textbf{h}_{r_pd}^k \hat{b}_{r_pd,k}(2i-1)+\textbf{h}_{r_qd}^k \hat{b}_{r_qd,k}(2i)+\textbf{n}(2i-1),$\\ \ \ \ \ \ \ \ \ \ \ \ \ \
$\textbf{y}_{r_{p,q}d,k}(2i)=\textbf{h}_{r_qd}^k \hat{b}^*_{r_pd,k}(2i-1)-\textbf{h}_{r_pd}^k \hat{b}^*_{r_qd,k}(2i)+\textbf{n}(2i).$\\  \ \ \ \ \ \ \ \ \
\%Apply the detectors/ML at the destination for detection\\ \ \ \ \ \ \ \ \ \ \ \ \
\textbf{break}\\ \ \ \ \ \ \ \
\textbf{else}\%choose the second highest SINR\\ \ \ \ \ \ \ \ \ \ \ \ \ \
${\rm SINR^{pre}_{p,q}}={\rm SINR_{p,q}}$\\ \ \ \ \ \ \ \ \ \ \ \ \ \
${\rm SINR_{p,q}} \in {\rm max} \{ {\rm SINR_{sr_{m,n}}},  {\rm SINR_{r_{m,n}d}}\} \setminus {\rm SINR^{pre}_{p,q}}$\\ \ \ \ \ \ \ \ \
\textbf{end}\\ \ \ \
\textbf{end}\\
\%Re-calculated the SINR for different link combinations and \\\
repeat the above process\\
\hline
\end{tabular}
\end{table}

\section*{Greedy Relay Pair Selection Technique}
In this section, a greedy relay pair selection algorithm is
introduced. For this relay selection problem
\cite{hesketh2014joint}, the exhaustive search of all possible relay
pairs is the optimum way to obtain the best performance. However,
the major problem that prevents us from applying this method when a
large number of relays involved in the transmission is its
considerable computational complexity. When $L$ relays ($L/2$ relay
pairs if $L$ is an even number) participate in the transmission, a
cost of $L(L-1)$ link combinations \cite{xu2015adaptive} is required
as both source-relay links and relay-destination links need to be
considered. Consequently, this fact motivates us to seek alternative
approaches that can achieve a good balance between performance and
complexity.

We propose a greedy relay pair selection algorithm that can approach the global
optimum with a reduced computational complexity. The algorithm starts with a
single link selection where we examine the SINR for each of the links as given
by
\begin{equation}
\scriptsize{{\rm SINR}_{sr_{p}}=\frac{\sum\limits_{k=1}^K \textbf{w}_{s_kr_p}^H \rho_{s_kr_p}\textbf{w}_{s_kr_p} }{\sum\limits_{k=1}^K \sum\limits_{\substack{l=1\\l\neq p}}^L \textbf{w}_{s_kr_l}^H \rho_{s_kr_l}\textbf{w}_{s_kr_l} + \sigma^2\textbf{w}_{s_kr_p}^H \textbf{w}_{s_kr_p}},}
\end{equation}
\begin{equation}
\scriptsize{{\rm SINR}_{r_{p}d}=\frac{\sum\limits_{k=1}^K (\textbf{w}_{r_pd}^k)^H \rho_{r_pd}^k \textbf{w}_{r_pd}^k}{\sum\limits_{k=1}^K \sum\limits_{\substack{l=1\\l\neq p}}^L (\textbf{w}_{r_ld}^k)^H \rho_{r_ld}^k \textbf{w}_{r_ld}^k + \sigma^2(\textbf{w}_{r_pd}^k)^H(\textbf{w}_{r_pd}^k)},}
\end{equation}
where ${\rm SINR}_{sr_{p}}$ and ${\rm SINR}_{r_{p}d}$ denote the SINR from the source to an arbitrary relay $p$ and from relay $p$ to the destination, respectively. We then select the link with the highest SINR and its associated relay $q$ is recorded as the base relay and given by
\begin{equation}
{\rm SINR^{base}_q}={\rm arg}\max _{\substack{p\in[1,2,...,L]}} \{{\rm SINR_{sr_p}}, {\rm SINR_{r_pd}}\}.
\end{equation}
Consequently, all possible relay pairs involved with base relay $q$
are listed as $\Omega_{p,q}$, where $p\in [1,L], p\neq q$. The SINR
for these $(L-1)$ relay pairs are then calculated as in
(\ref{equation3}) and (\ref{equation4}). After that, the optimum
relay pair $\Omega_{n,q}$ is chosen according to (\ref{equation5})
and the algorithm begins if the corresponding buffers are available
for either transmission or reception.
\\
\\
\textbf{Transmission mode:}\\
When the buffers are switched to the transmission mode, a buffer space check is
conducted firstly to ensure the corresponding buffers are not empty. We then
have,
\begin{equation}
\Omega^{\rm buffer}_{n}\neq\varnothing, \  n\in [1,2,...,L],
\end{equation}
and
\begin{equation}
\Omega^{\rm buffer}_{q}\neq\varnothing, \ q\in [1,2,...,L],
\end{equation}
where $\Omega^{\rm buffer}_{n}$ and $\Omega^{\rm buffer}_{q}$ represents the
buffer $n$ and the buffer $q$ associated with the relay pair $\Omega_{n,q}$. In
this situation, the DSTC scheme is performed afterwards as in
(\ref{equation10}) and (\ref{equation11}) through the selected relay pair.
Conversely, empty buffer entries indicate that the selected relay pair is not
capable of forwarding the data to the destination. In this case, we drop this
relay pair, select another relay pair among the remaining $(L-2)$ candidate
pairs with the highest SINR as given by
\begin{equation}\label{equation12}
{\rm SINR_{m,q}} = {\rm arg}\max _{\substack{p\neq n,q\\p \in [1,2,...,L]}} \{  {\rm SINR_{p,q}}  \},
\end{equation}
The algorithm then repeats the new selected relay pair $\Omega_{m,q}$.
Otherwise, if all possible relay pairs $\Omega_{p,q}$ ($p\neq n,q, p \in
[1,L]$) are not available, we then reset the base relay associated with the
second highest SINR as described by
\begin{equation}\label{equation13}
{\rm SINR^{pre}_{q}}={\rm SINR^{base}_q},
\end{equation}
\begin{equation}\label{equation14}
{\rm SINR^{base}_q} = {\rm max} \{{\rm SINR_{sr_p}}, {\rm SINR_{r_pd}} \} \setminus {\rm SINR^{pre}_{q}},
\end{equation}
where $\{{\rm SINR_{sr_p}}, {\rm SINR_{r_pd}} \} \setminus {\rm
SINR^{pre}_{q}}$ denotes a complementary set where we drop
the ${\rm SINR^{pre}_{q}}$ from the link SINR set $\{{\rm SINR_{sr_p}}, {\rm SINR_{r_pd}} \}$.
After this selection process, a new relay pair is chosen and the transmission procedure repeats as above according to the buffer status.\\
\\
\textbf{Reception mode:}\\
When the buffers are switched to reception mode, similarly, a buffer space
check is performed initially to ensure there is enough space for storing the
processed data, namely,
\begin{equation}
\Omega^{\rm buffer}_{n}\neq {\rm U}, \ n\in [1,2,...,L],
\end{equation}
and
\begin{equation}
\Omega^{\rm buffer}_{q}\neq {\rm U}, \ q\in [1,2,...,L],
\end{equation}
where ${\rm U}$ represents a full buffer set. In this case, if the buffers are
not full, then, the sources send the data to the selected relay pair
$\Omega_{n,q}$ over two time instants according to (\ref{equation6}) and
(\ref{equation7}). Otherwise, the algorithm reselects a new relay pair as in
(\ref{equation12}), (\ref{equation13}) and (\ref{equation14}).

{ In summary, the relay pair selection algorithm solves a
combinatorial problem using exhaustive searches by comparing the
SINR of all links and combinations. Alternatively, a low-complexity
algorithm (for example, the proposed greedy algorithm) could be used
to reduce the computational complexity of the pair selection task. }

The greedy relay pair selection algorithm is show in Table 2.

\begin{table}[!htp]
\centering\caption{The proposed greedy relay pair selection algorithm}
\begin{tabular}{l}
\hline
\%Choose a single relay with the highest SINR that \\ \ \ \ \
corresponds to a specific base relay $q$\\ \ \ \ \
${\rm SINR^{base}_q}={\rm max} \{{\rm SINR_{sr_p}}, {\rm SINR_{r_pd}}\}, p\in [1,L]$\\ \ \ \ \
\textbf{For} ${\rm p=1:L}$ \% all relay pairs associated with relay $q$\\ \ \ \ \ \ \ \ \ \ \ \
\textbf{if} ${\rm p\neq q}$ \\ \ \ \ \ \ \ \ \ \ \ \ \ \ \
$\Omega_{\rm relay pair}={\rm [p,q]}$\\ \ \ \ \ \ \ \ \ \ \
\% when the links belong to the source-relay phase\\ \ \ \ \ \ \ \ \ \ \
$\scriptsize{{\rm SINR}_{sr_{p,q}}=\frac{\sum\limits_{k=1}^K \textbf{w}_{s_kr_p}^H \rho_{s_kr_p} \textbf{w}_{s_kr_p}+\textbf{w}_{s_kr_q}^H \rho_{s_kr_q} \textbf{w}_{s_kr_q} }{\sum\limits_{k=1}^K \sum\limits_{\substack{l=1\\l\neq p,q}}^L \textbf{w}_{s_kr_l}^H \rho_{s_kr_l} \textbf{w}_{s_kr_l} + \sigma^2\textbf{w}_{s_kr_p}^H \textbf{w}_{s_kr_p}+ \sigma^2\textbf{w}_{s_kr_q}^H \textbf{w}_{s_kr_q}},}$\\ \ \ \ \ \ \ \ \ \ \
\% when the links belong to the relay-destination phase\\ \ \ \ \ \ \ \ \ \ \
$\scriptsize{{\rm SINR}_{r_{p,q}d}=\frac{\sum\limits_{k=1}^K (\textbf{w}_{r_pd}^k)^H \rho_{r_pd}^k \textbf{w}_{r_pd}^k+(\textbf{w}_{r_qd}^k)^H \rho_{r_qd}^k \textbf{w}_{r_qd}^k}{\sum\limits_{k=1}^K \sum\limits_{\substack{l=1\\l\neq p,q}}^L (\textbf{w}_{r_ld}^k)^H \rho_{r_ld}^k \textbf{w}_{r_ld}^k + \sigma^2(\textbf{w}_{r_pd}^k)^H \textbf{w}_{r_pd}^k+ \sigma^2(\textbf{w}_{r_qd}^k)^H \textbf{w}_{r_qd}^k},}$\\ \ \ \ \ \ \ \ \ \ \
\% record each calculated relay pair SINR\\ \ \ \ \ \ \ \ \ \
\textbf{end}\\ \ \ \ \
\textbf{end}\\ \ \ \ \
${\rm SINR_{n,q}}= {\rm max} \{ {\rm SINR}_{sr_{p,q}} \ , \ {\rm SINR}_{r_{p,q}d}  \}$\\ \ \ \ \
\textbf{if}\%Reception mode\\ \ \ \ \ \ \ \
\textbf{if} the buffers entries are not full\\ \ \ \ \ \ \ \ \ \ \
$\textbf{y}_{sr_{n,q}}(2i-1)=\sum\limits_{k=1}^K \textbf{h}_{s_kr_{n,q}} b_k(2i-1)+\textbf{n}(2i-1),$\\ \ \ \ \ \ \  \ \ \ \
$\textbf{y}_{sr_{n,q}}(2i)=\sum\limits_{k=1}^K \textbf{h}_{s_kr_{n,q}}b_k(2i)+\textbf{n}(2i).$\\ \ \ \ \ \ \
\%Apply the detectors at relay $n$ and relay $q$ to obtain\\\ \ \ \ \ \ \ \ \ \
 $\hat{b}_{r_{n,q}d,k}(2i-1)$ and $\hat{b}_{r_{n,q}d,k}(2i)$ and store them \\ \ \ \ \ \ \ \ \ \ \
in the corresponding buffer entries\\ \ \ \ \ \
\textbf{ else} \%choose another link with the second highest SINR\\ \ \ \ \ \ \ \ \ \ \ \
${\rm SINR^{pre}_{q}}={\rm SINR^{base}_q}$\\ \ \ \ \ \ \ \ \ \ \ \
${\rm SINR^{base}_q} \in {\rm max} \{{\rm SINR_{sr_p}}, {\rm SINR_{r_pd}} \} \setminus {\rm SINR^{pre}_{q}}$\\ \ \ \ \ \ \ \
\%Repeat the above greedy relay pair selection process\\ \ \ \ \ \ \
\textbf{end}\\ \ \
\textbf{else}\%Transmission mode\\ \ \ \ \ \ \
\textbf{if} the buffers entries are not empty\\ \ \ \ \ \ \ \ \ \
$\textbf{y}_{r_{n,q}d,k}(2i-1)= \textbf{h}_{r_nd}^k \hat{b}_{r_nd,k}(2i-1)+\textbf{h}_{r_qd}^k \hat{b}_{r_qd,k}(2i)+\textbf{n}(2i-1),$\\ \ \ \ \ \ \ \ \ \
$\textbf{y}_{r_{n,q}d,k}(2i)= \textbf{h}_{r_qd}^k \hat{b}^*_{r_nd,k}(2i-1)-\textbf{h}_{r_nd}^k \hat{b}^*_{r_qd,k}(2i)+\textbf{n}(2i).$\\  \ \ \ \ \ \ \ \
\%Apply the detectors/ML at the destination for detection\\ \ \ \ \ \
\textbf{else}\%choose another link with the second highest SINR\\ \ \ \ \ \ \ \ \ \ \ \ \
${\rm SINR^{pre}_{q}}={\rm SINR^{base}_q}$\\ \ \ \ \ \ \ \ \ \ \ \ \
${\rm SINR^{base}_q} \in {\rm max} \{{\rm SINR_{sr_p}}, {\rm SINR_{r_pd}} \} \setminus {\rm SINR^{pre}_{q}}$\\ \ \ \ \ \ \ \ \ \
\%Repeat the above greedy relay pair selection process\\ \ \ \ \ \
\textbf{end}\\ \ \ \
\textbf{end}\\
\%Repeat the above greedy relay pair selection process\\
\hline
\end{tabular}
\end{table}

\section*{Proposed Dynamic Buffer Scheme}
The size $J$ of the buffers also plays a key role in the performance
of the system, which improves with the increase of the size as buffers with
greater size allow more data packets to be stored. In this case, extra degrees
of freedom in the system or choices for data transmission are available. Hence, in this section, we release the limitation on the size of the buffer to further explore the additional advantage brought by dynamic buffer design where the buffer size can vary according to different criteria such as the input SNR and the channel condition.
When considering the input SNR, larger buffer space is required when the transmission is operated in low SNR region so that the most proper data can be selected among a greater number of candidates. On the other hand, in the high SNR region, a small buffer size is employed as most of the processed symbols are appropriate when compared with the situation in the low SNR region. In this work, we assume that the buffer size $J$ is inversely proportional to the input SNR, namely, with the increase of the SNR, the buffer size decreases automatically. The algorithm for calculating the buffer size $J$ is detailed in Table. \ref{table3}.
\begin{table}[!htb]
\centering\caption{The algorithm to calculate the buffer size $J$}
\begin{tabular}{l}
\hline
\textbf{If}\ \ \ \ \ \ \ ${\rm SNR_{cur}}={\rm SNR_{pre}}+d_1$ \\
\\
\textbf{then} \ \ \ \  $ J_{\rm cur}=J_{\rm pre}-d_2$,\\
\\
where ${\rm SNR_{cur}}$ and ${\rm SNR_{pre}}$ represent the input SNR after and before\\
increasing its value,\\
\\
$J_{\rm cur}$ and $J_{\rm pre}$ denote the corresponding buffer size before and after\\
decreasing its value,\\
\\
$d_1$ and $d_2$ are the step sizes for the SNR and the buffer size, respectively.\\
\hline
\end{tabular}\label{table3}
\end{table}\\

The buffer size can be determined by the current selected channel pair condition. In particular, we set a threshold $\gamma$ that denotes the channel power, if the current selected channel power is under $\gamma$, the buffer size increases as more candidates need to be saved in order to select the best symbol, on the contrary, if the current selected channel pair power exceeds $\gamma$, we decrease the buffer size as there is a high possibility that the transmission is not significantly affected. The approach based on the channel power for varying the buffer can be summarized in Table. \ref{table4}.
\begin{table}[!htb]
\centering\caption{The algorithm for calculate buffer size $J$ based on the channel power}
\begin{tabular}{l}
\hline
\textbf{If}\ \ \ \ \ \ $\min\| h_{s_kr_l}\|^2 \leq \gamma$ \ \ \  \textbf{or} \ \ \ $\min\| h_{r_ld}\|^2 \leq \gamma, \ \ l\in[1,L]$ \\
\ \ \ \ \ \ \ \ \ $J_{\rm cur}=J_{\rm pre}+d_3$ \\
\textbf{else}\\
\ \ \ \ \ \ \ \ \ $J_{\rm cur}=J_{\rm pre}-d_3$\\
\textbf{end}\\
\\
where $d_3$ represents the step size when adjusting the buffer size.\\
\hline
\end{tabular}\label{table4}
\end{table}\\

\section*{Analysis of the Proposed Algorithms}
In this section, we analyse the computational complexity required by the proposed relay pair selection algorithm, the problem of the average delay brought by the proposed schemes and algorithms, followed by the discussion of the proposed greedy algorithm.

\subsection*{Computational Complexity}
The proposed greedy relay pair selection method considers the
combination effect of the channel condition so that the DSTC
algorithm can be applied with a collection of relays. When compared
with the exhaustive search that lists all possible subsets of relay
pairs, less than $L(L-1)$ types of link combinations { (associated
with the corresponding $L(L-1)/2$ relay pairs)} are examined as the
proposed method explores the link combination when both single relay
and relay pair are involved. For the greedy relay selection
strategy, the proposed scheme explores a moderate to large number of
relay pairs at each stage, however, the algorithm stops when the
corresponding entries satisfy the current system requirement
(transmission mode or reception mode), in this case, the maximum
number of relay pair that we have to examine is
$(L-1)+(L-2)+...+1=L(L-1)/2$. On the other hand, when consider the
exhaustive search, the total number of relay pairs that must be
verified is { $C_L^2=L(L-1)/2$}. It should be mentioned that when
calculating the associated SINR, we have to double the number of
calculation flops as we have to consider and compare both the SINR
for source-relay links and relay-destination links. The detailed
computational complexity is listed in Table. \ref{table5}. When
compare these two algorithms, the proposed greedy relay pair
selection algorithm is an order of magnitude less costly. The
complexly of channel estimation and receive filter computation are
also listed in Table. \ref{table5}. It can been seen that, when a
large number of relays participate in the transmission, with a
careful control of the buffer size $J$, a good balance of complexity
and performance is achieved.
\begin{table}[!htb]
\centering\caption{Computational complexity}
\begin{tabular}{l|c|c|c}
\hline
Processing&Algorithm&Multiplications&Additions\\
\hline
Relay pair&Exhaustive&$7KNL^3$&$(2KN+K)L^3+2L$\\
selection&Search&$-7KNL^2$&$-(2KN+K+2)L^2$\\
\hline
Relay pair&Greedy&$21KNL^2$&$6KNL^2+3KL^2$ \\
selection&Search&$-7KNL$&$-3KL-L+1$\\
\hline
Channel&Exhaustive Search&$(2N+1)KL$&$(2N-1)KL$\\
estimation&Greedy Search&&\\
\hline
Receive filter&Exhaustive Search&$4NJ$&$(4N-2)J$\\
computation&Greedy Search&&\\
(RAKE)&&\\
\hline
\end{tabular}\label{table5}
\end{table}\\

\subsection*{Average Delay Analysis}
The improvement of the performance brought by the buffer-aided relays comes at the expense of the transmission delay. Hence, it is of great importance to investigate the performance-delay trade-off of the proposed buffer-aided DSTC schemes \cite{Islam}. In this subsection, we analyze the average delay of the proposed schemes and algorithms.

We assume that the source always has data to transmit and the delay is mostly caused by the buffers that equip the relays. Let $T(i)$ and $Q(i)$ denote the delay of packets of $M$ symbols transmitted by the source and the queue length at time instant $i$ for DSTC schemes, respectively.

According to Little's law \cite{Bertsekas}, the average delay, which is also the average time that packets are stored in the corresponding buffer is given by
\begin{equation}
T=\frac{Q}{R_a} \ \ \ \ {\rm time \ \ slots,}
\end{equation}
where $Q=E[Q(i)]$ represents the average queue length at the relay buffer, $R_a$ (in packets/slot) is the average arrival rate into the queue.

In this analysis, we assume both the source and relay transmit at a constant instantaneous rate $R$ ($R=1 \ \ {\rm packets/slot}= M \ \ {\rm symbols/slot}$) when they are selected for transmission and the transmission is operated with one packets of $M$ symbols per each time slot. We also for simplicity define the error probability for the source-relay link and relay-destination link as $P_{sr}$ and $P_{rd}$ ($P=P_{sr}=P_{rd}$), respectively. For a buffer with size $J$ ($J$ packets are stored in the buffer), the average queue length is described by \cite{Islam}
\begin{equation}
Q=\sum_{j=0}^J jP_{G_j}=JP_{G_J},
\end{equation}
where $P_{G_j}$ represents the buffer state probability that has been explained in \cite{Islam}, $P_{G_J}=P_{\rm full}$ denotes the probability when the buffer is full. Similarly, we then define $P_{G_0}=P_{\rm empty}$ as the probability for empty buffer. Therefore, the average arrival rate into the buffer can be calculated as
\begin{equation}
R_a=(1-P_{G_J})P+P_{G_0}P,
\end{equation}
Similarly, the average departure rate from the buffer is given by as
\begin{equation}
R_d=(1-P_{G_0})P+P_{G_J}P.
\end{equation}
Consequently, the above equations can be further derived as
\begin{equation}
T=\frac{Q}{R_a}=\frac{P_{G_J}}{(1-P_{G_J})P+P_{G_0}P}J {\rm \ \ packets/slot}
\end{equation}
Clearly, the above results demonstrate that the transmission delay is linear with the buffer size.

Apart from that, the DSTC scheme will introduce further delay. For the DSTC scheme, the relay pair need to wait an extra time slot for the second packets to arrive. Then, the relay pair can transmit the packets to the destination using DSTC scheme. In other word, the DSTC scheme takes two time--slots to transmit two packets, as a result, it brings extra delay obviously \cite{Bouanen,Gong,Sheng}. Meanwhile, the relay pair selection processing also brings delay. For both exhaustive and greedy selection, they need to calculate the best relay pair from the candidates pool. This processing need extra computation time until the best relay pair is selected.

\subsection*{Greedy Relay Selection Analysis}
The proposed greedy relay pair selection method is a stepwise forward selection algorithm, where we optimize the selection based
on the SINR criterion at each stage. We begin the process with a single link selection where we examine the SINR for each of the links and choose the link with the highest SINR, the associated relay is then selected and the candidate relay pair is generated by adding the remaining relays, respectively. The optimum relay pair is subsequently selected according to the SINR criterion. Since buffers are equipped at each relay, it is possible that the corresponding relay pair entries are not available for either transmission or reception. In this case, the candidate relay pair from the first step with the second highest SINR is then chosen. Clearly, if all remaining candidate relay pairs are not selected due to the unavailability of the associated buffers, we reset the base relay and newly generated relay pairs are grouped in the second stage by adding other relays,  respectively. Obviously, the number of all possible relay pair candidates at each stage is reduced gradually as the discarded relay pair from previous stages will not appear in the current stage. Hence, the relay pairs grouped at each step are presented as follows:
\begin{equation*}
\begin{split}
&{\rm Stage \ 1:} \ \ \{\Omega_1^1, \Omega_2^1,...,\Omega_{L-1}^1\},\\
&{\rm Stage \ 2:} \ \ \{\Omega_1^2,\Omega_2^2,...,
\Omega_{L-2}^2\},\\
&\ \ \ \ \ \ \ \ \ \ \ \ \ \ \ \ \ \vdots \\
&{\rm Stage \ s:} \ \ \{\Omega_1^s,\Omega_2^s,...,\Omega_{L-s}^s\},\\
&\ \ \ \ \ \ \ \ \ \ \ \ \ \ \ \ \ \vdots \\
&{\rm Stage \ L-1:} \ \ \{\Omega_1^1 \},\\
\end{split}
\end{equation*}
where $\Omega_i^s$ denotes the $i$-th relay pair at the $s$-th stage. Clearly, the maximum number of relay pairs that we have to consider for all $L-1$ stages is $(L-1)+(L-2)+...+1=L(L-1)/2$, since this algorithm stops when selected relay pair with its associated buffers are available, the associated complexity for the proposed greedy relay selection strategy is less than $L(L-1)/2$.

Compared with the exhaustive search, which is considered as the optimum relay selection method, the number of relay pairs examined for the processing is given by
\begin{equation*}
{\rm Stage \ 1:} \ \ \{\Omega_1^1,\Omega_2^1,...,\Omega_{\frac {L(L-1)}{2}}^{1}\},\\
\end{equation*}
The total number of relay combinations can then be calculated as $C_L^2=L(L-1)/2$, where each term $C_m^n=\frac{m(m-1)...(m-n+1)}{n!}$ represents the number of combinations that we choose, i.e., $n$ elements from $m$ elements$(m\geq n)$.

Because the number of relay pairs that we have to consider for the greedy algorithm is less than exhaustive search, the proposed greedy algorithm provides a much lower cost in terms of flops and running time when compared with the exhaustive search. In fact, the idea behind the proposed algorithm is to choose relay pairs in a greedy fashion. At each stage, we select the set of relays with the highest SINR. Then we consider the availability of the buffers, if the corresponding buffer entries do not satisfy the system mode, we reselect the relay pair in the following stages. After several stages, the algorithm is able to identify the optimum (or a near optimum) relay set that can satisfy the current transmission. To this end, we state the following proposition. \\
\\
\textit{Proposition}: the proposed greedy algorithm achieves an SINR that is upper bounded as follows:
\begin{equation}
\begin{split}
{\rm SINR_{\Omega greedy}} \leq {\rm SINR_{\Omega exhaustive}}
\end{split}
\end{equation}
Proof:



We investigate the upper bound by comparing the proposed algorithm and the exhaustive search at the first stage. At stage 1, since $\Omega_{\rm greedy}^s$ is a candidate subset of the exhaustive search, we have
\begin{equation}
\Omega_{\rm exhaustive}^1={\rm max} \ \{ \Omega_{\rm exhautive(i)}^{1}, i\in[1,C_L^{2}] \},
\end{equation}
\begin{equation}
\Omega_{\rm greedy}^s \in \{ \Omega_{\rm exhautive(i)}^{1}, i\in[1,C_L^{2}]\},
\end{equation}
where $\Omega_{\rm exhaustive(i)}^1$ represents the $i$-th relay pair selected at the 1st stage of the exhaustive relay selection method.

Assuming both strategies select the same relay pair and the greedy algorithm is conducted at stage $s$, we have
\begin{equation*}
\begin{split}
&\Omega_{\rm greedy}^s= \ \{p,q\},\\
&\Omega_{\rm exhaustive}^1= \ \{p,q\},\\
\end{split}
\end{equation*}
this situation again leads to the equality that ${\rm SINR}_{\rm \Omega_{greedy}^s}={\rm SINR}_{\rm \Omega_{exhaustive}^1}$. In contrast, if the exhaustive search chooses another relay set that belongs to $\{ \Omega_{\rm exhautive(i)}^{1}, i\in[1,C_L^{2}] \}$ that provides a higher SINR, clearly, $\Omega_{\rm greedy}^s \neq \Omega_{\rm exhaustive}^1$, we can then obtain the inequality that ${\rm SINR}_{\Omega_{\rm greedy}^s} \leq {\rm SINR}_{\Omega_{\rm exhaustive}^1}$.

\section*{simulations}

In this section, a simulation study of the proposed buffer-aided
DSTC techniques for cooperative systems is carried out. The DS-CDMA
network uses randomly generated spreading codes of length $N=16$.
The corresponding channel coefficients are modeled as uniformly
random variables and are normalized to ensure the mean
signal value over all transmissions is unity for all analyzed techniques.
We assume perfectly known channels at the receivers and we also present
an example with channel estimation. Equal power allocation is employed.
We consider packets with 1000 BPSK symbols and step size $d=2$ when
evaluating the dynamic schemes. We consider fixed buffer--aided
exhaustive/greedy (FBAE/FBAG) relay pair selection strategies (RPS)
and dynamic buffer--aided exhaustive/greedy (DBAE/DBAG) RPS. 



In order to verify that the fixed buffer-aided relay pair DSTC
cooperative scheme contributes to the performance gain, we compare
the performance between the situations of the transmission with
fixed size buffers and without buffers in Fig. 3. The first example
shown in Fig. 3(a) illustrates the performance comparison between
the proposed buffer-aided DSTC transmission with different RPS and
DSTC transmission with different RPS and no buffers when better
decoding techniques are adopted. The system has 3 users, 6 relays,
perfect decoding is assumed  at each relay and the matched filter is
adopted at the destination. Specifically, for the no relay selection
(RS) DSTC technique, all relays participate in the DSTC transmission
(every two consecutive relays are working in pairs). Similarly, for
the non buffer-aided schemes, the RPS process only occurs during the
second phase (relay-destination), where the random selection
algorithm chooses an arbitrary relay pair, the proposed greedy
algorithm chooses two relays associated with two optimum
relay-destination links and the exhaustive relay pair schemes
examines all possible relay pairs and selects the one with the
highest SINR. In contrast, the proposed buffer-aided scheme
automatically selects the relay pair over both source-relay links
and relay-destination links. Moreover, with the help of the buffers,
the most appropriate data are sent and better overall system
performance can be achieved. { As for different decoding methods, we
have also evaluated the BER performance when the ML detector is
applied at the destination and the results show that ML detector
significantly outperforms the simple RAKE receiver, as expected.}
Apart from that, the performance for a single-user buffer-aided
exhaustive RPS DSTC is presented here for comparison purposes.
Consequently, the results reveal that our proposed buffer-aided
strategies ($J=6$) perform better than the one without buffers. In
particular, Fig. 3(a) also illustrates that our proposed
buffer-aided schemes can approach the single-user bound very
closely.

{  Another example depicted in Fig. 3(b) compares the proposed
buffer-aided DSTC transmission with different RPS and non-buffer
aided DSTC transmission with different RPS}. In this scenario, where
we apply the linear MMSE receiver at each of the relay and the RAKE
at the destination in an uplink cooperative scenario with 3 users, 6
relays and buffer size $J=6$. {  Similarly, the system gain brought
by the use of the ML detector at the destination and the performance
bounds for a single-user buffer-aided exhaustive RPS DSTC are
presented for comparison purposes. The results also indicate that
the proposed buffer-aided strategies ($J=6$) have the highest
diversity gain when compared with the ones without buffers.}
Furthermore, the BER performance curves of the greedy RPS algorithm
approaches the exhaustive RPS, while keeping the complexity
reasonably low for practical use.


In the second example, we compare the proposed buffer-aided DSTC
transmission with different RPS strategies and DSTC transmission
with RPS and no buffers with channel estimation. The results are
shown in Fig. 4. In this scenario, where we apply the linear MMSE
receiver at each of the relay and the RAKE at the destination in an
uplink cooperative scenario with 3 users, 6 relays and buffer size
$J=6$. Clearly, it can been seen that, due to the introduction of
channel estimation, the performance for all algorithms are slightly
degraded when compared with the assumption of perfect CSI. However,
{ our proposed buffer-aided strategies ($J=6$) still perform better
than the one without buffers when referring to the diversity gain.}


The third example illustrates the performance comparison for the
fixed buffer-aided design in Fig. 5(a) and dynamic buffer-aided
design in Fig. 5(b) in a cooperative DSTC system with different
relay pair selection strategies (RPS). The overall network has 3
users, 6 relays, the linear MMSE receiver is applied at each relay
and the RAKE receiver is adopted at the destination. For dynamic
algorithms, the buffer size $J$ decreases when approaching higher
SNR region. In both figures, the buffer-aided exhaustive RPS
algorithm performs better than the greedy one. When we compare the
two figures in Fig. 5, the dynamic buffer techniques are more
flexible than the fixed buffer ones as they explore the most
suitable buffer size for the current transmission according to a
given criterion. In this case, there is a greater possibility to
select the most appropriate data when the transmission is operated
in poor condition as more candidates are stored in the buffer space.
On the other hand, the transmission delay can be avoided when the
outer condition improves as most of the candidates are appropriate.
Simulation results verify these points and indicate that the
DBAE/DBAG RPS outperform the FBAE/FBAG ($J=8$) RPS and the advantage
increases when adopting the single user case. Furthermore, it can
also be seen that the BER performance curves of the greedy relay
pair selection algorithm \cite{jiaqi,gu2016joint} approaches the
exhaustive search, whilst keeping the complexity reasonably low for
practical utilization.


The fourth example compares the FBAE/FBAG RPS scheme in Fig. 6(a)
and the DBAE/DBAG RPS strategy in Fig. 6(b) in a DSTC cooperative
system, where we apply the linear MMSE receiver at each of the relay
and the RAKE receiver at the destination in an uplink cooperative
scenario with 3 users, 6 relays and fixed buffer size $J=8$.
Similarly, the performance for a single-user buffer-aided exhaustive
RPS DSTC is presented for comparison purposes. In both figures, the
buffer-aided exhaustive search RPS algorithm performs better than
the greedy one. The average dynamic buffer size $J$ is highly
dependant on the threshold $\gamma$ and the step size $d$, clearly,
with careful control on these parameters, better performance can be
achieved. The simulation results also indicate that the proposed
dynamic design perform better than the fixed buffer size ones when
we apply the same relay selection method, as depicted in Fig. 6.


The algorithms are then assessed in terms of the BER versus buffer
size $J$ in Fig. 7 with a fixed SNR=15dB. In this scenario, we
assume perfect decoding at the relays as accurate detection at
relays would highly influence the following transmission and apply
the RAKE at the destination. The results indicate that the overall
BER degrades as the size of the buffer increases. It also shows that
with larger buffer sizes, the system experiences diminishing returns
in performance. In this case, a good balance between the
transmission delay and the buffer size can be obtained when the
buffer size is carefully considered.


In the last part, we demonstrate the influence of the dynamic buffer
size on the average delay. We examine via simulations the proposed
algorithms by measuring the average delay in packets versus the
packet size in symbols in Fig. 8. This figure presents the average
delay for different algorithms, when a certain number of symbols
(that form packets) are transmitted. For the average delay, we
measure the number of extra packets that are employed when compared
with transmitted symbols. There exists a linear relation between the
average delay and transmitted symbols, as outlined in the delay
analysis. This is because for each certain number of transmitted
packets, the average delay is similar. As a result, when the
transmitted symbols increase with a fixed numbers of packets, the
average delay also increases with a similar numbers of packets.
Therefore, with the increase of transmitted packet size, the average
delay increases together. When we compare the proposed algorithms,
the dynamic buffer size reduces the average delay. In particular,
the proposed multiuser buffer--aided greedy DSTC algorithm with
dynamic buffer scheme has the lowest delay, followed by the greedy
DSTC algorithm, the exhaustive DSTC algorithm with and without the
dynamic buffer scheme.

\section*{conclusions}

In this work, we have presented a dynamic buffer-aided DSTC scheme
for cooperative DS-CDMA systems with different relay pair selection
techniques. With the help of the dynamic buffers, this approach
effectively improves the transmission performance and help to
achieve a good balance between bit error rate (BER) and delay. We
have developed algorithms for relay-pair selection based on an
exhaustive search and on a greedy approach. A dynamic buffer design
has also been devised to improve the performance of buffer-aided
schemes. Simulation results show that the performance of the
proposed scheme and algorithms can offer good gains as compared to
previously reported techniques.

\begin{backmatter}
\section*{Competing interests}
The authors declare that they have no competing interests.

\section*{Acknowledgements}
This work is funded by the ESII consortium under task 26 for low-cost wireless ad hoc and sensor networks.

\section*{Figure legends}
Figure 1: Uplink of a cooperative DS-CDMA system.

Figure 2: Proposed buffer-aided cooperative scheme.

Figure 3(a):Performance comparison for buffer-aided scheme and non buffer-aided scheme in cooperative DS-CDMA system with perfect decoding at the relay, RAKE at the destination.

Figure 3(b):Performance comparison for buffer-aided scheme and non buffer-aided scheme in cooperative DS-CDMA system with MMSE at the relay, RAKE at the destination.

Figure 4: Performance comparison for buffer-aided scheme and non buffer-aided scheme in cooperative DS-CDMA system with MMSE at the relay, RAKE at the destination with channel estimation applied.

Figure 5(a): Performance comparison for fixed buffer design (input SNR criterion).

Figure 5(b): Performance comparison for dynamic buffer design (input SNR criterion).

Figure 6(a): Performance comparison for fixed buffer design (channel power criterion).

Figure 6(b): Performance comparison for dynamic buffer design (channel power criterion).

Figure 7: BER versus size of the buffers for uplink cooperative system.

Figure 8: Packet size comparison for uplink cooperative system.

\bibliographystyle{bmc-mathphys}
\bibliography{reference}

\end{backmatter}
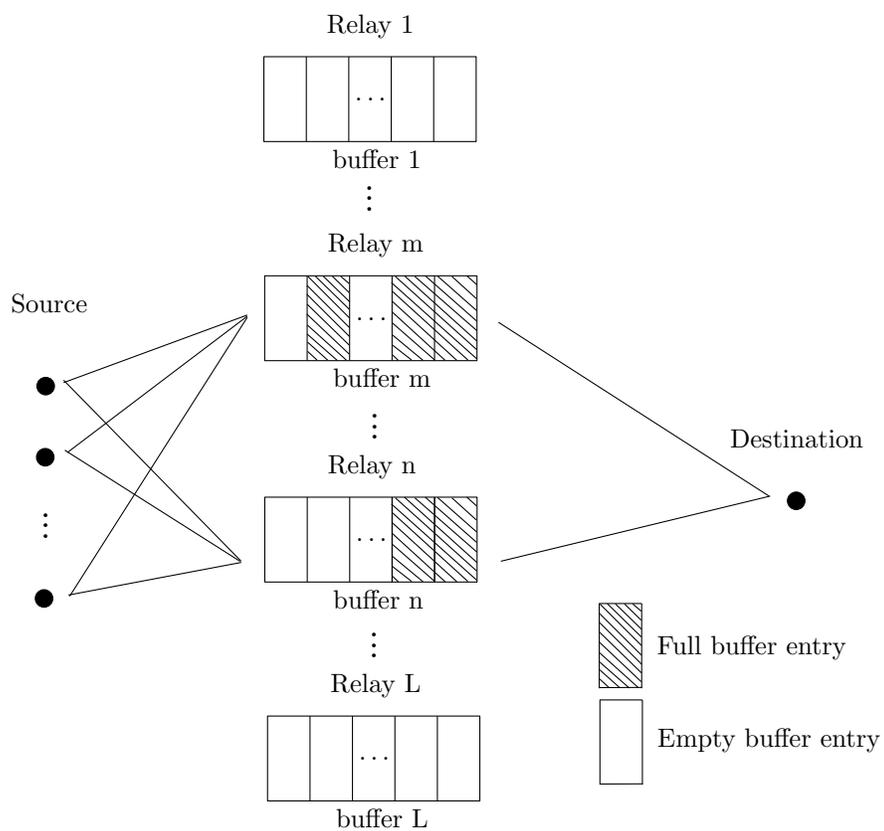
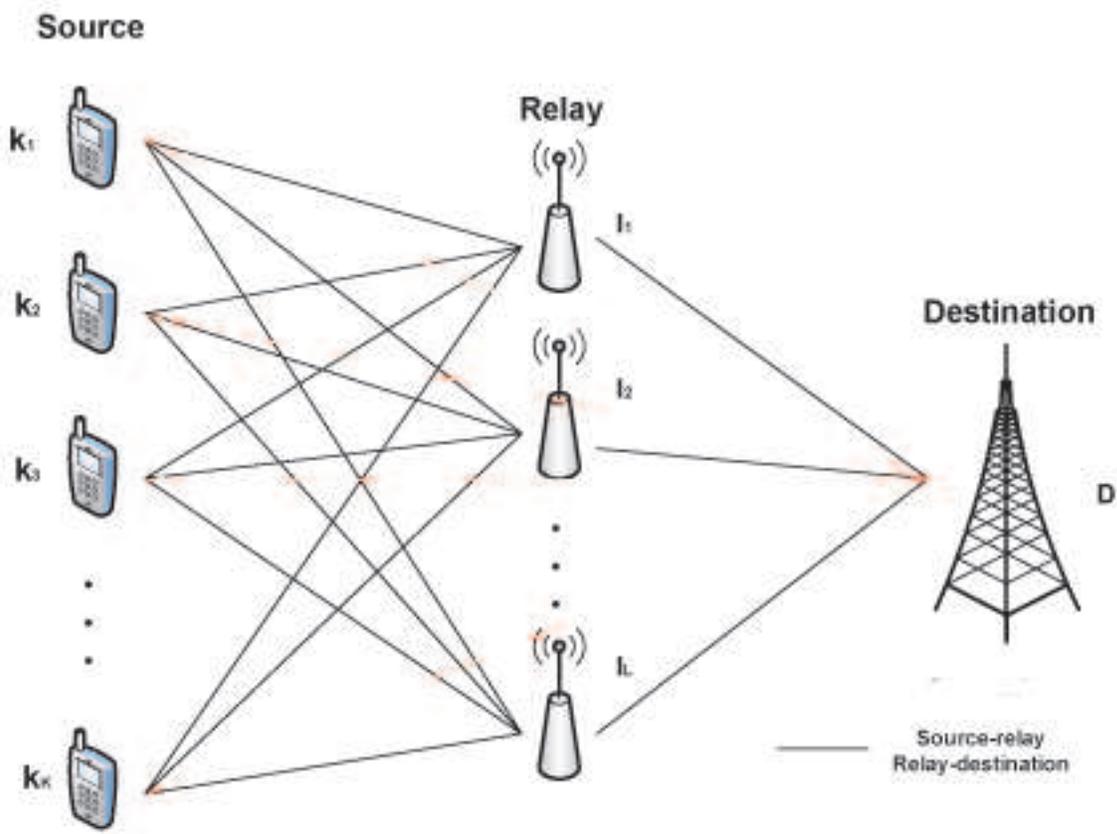
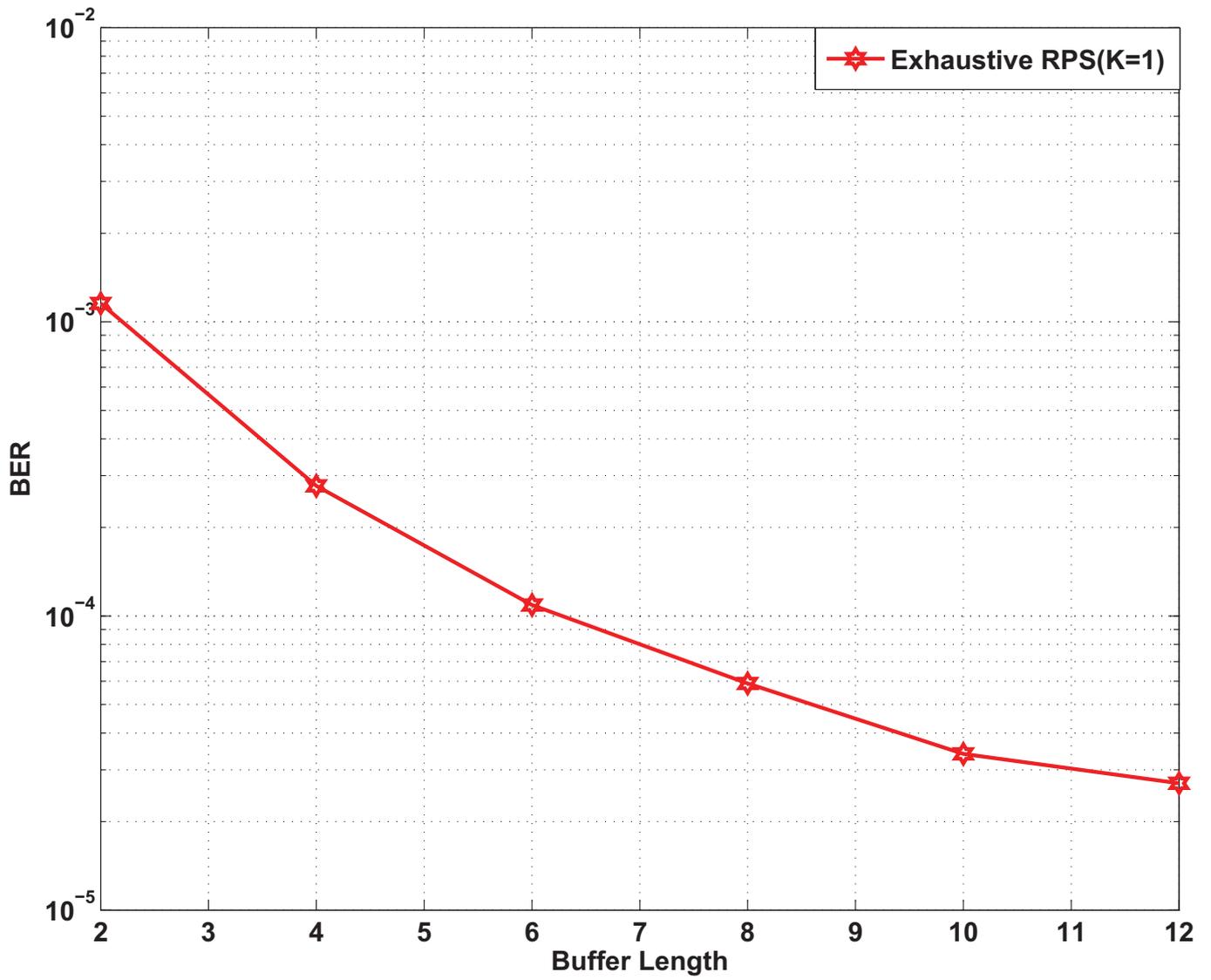
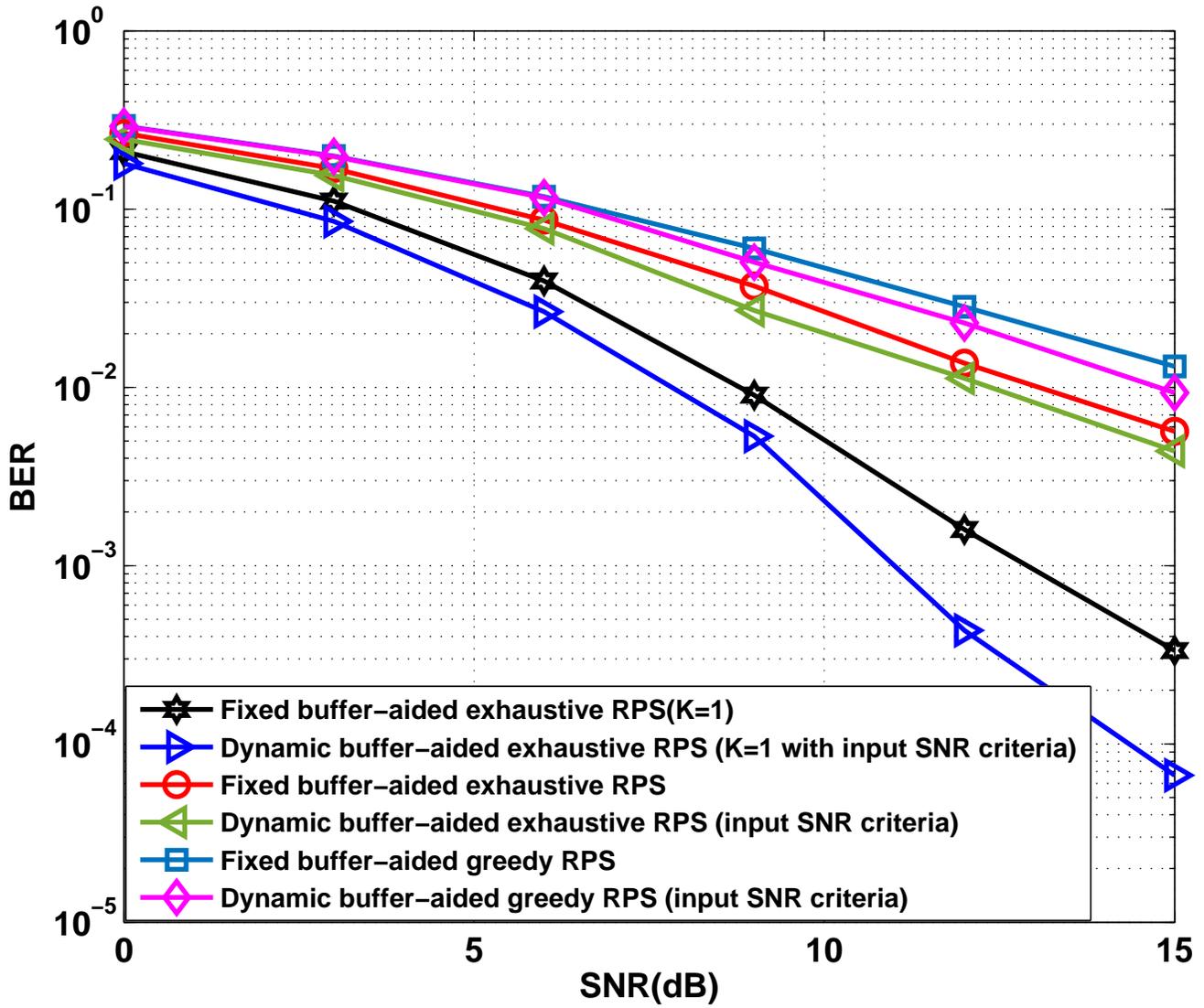
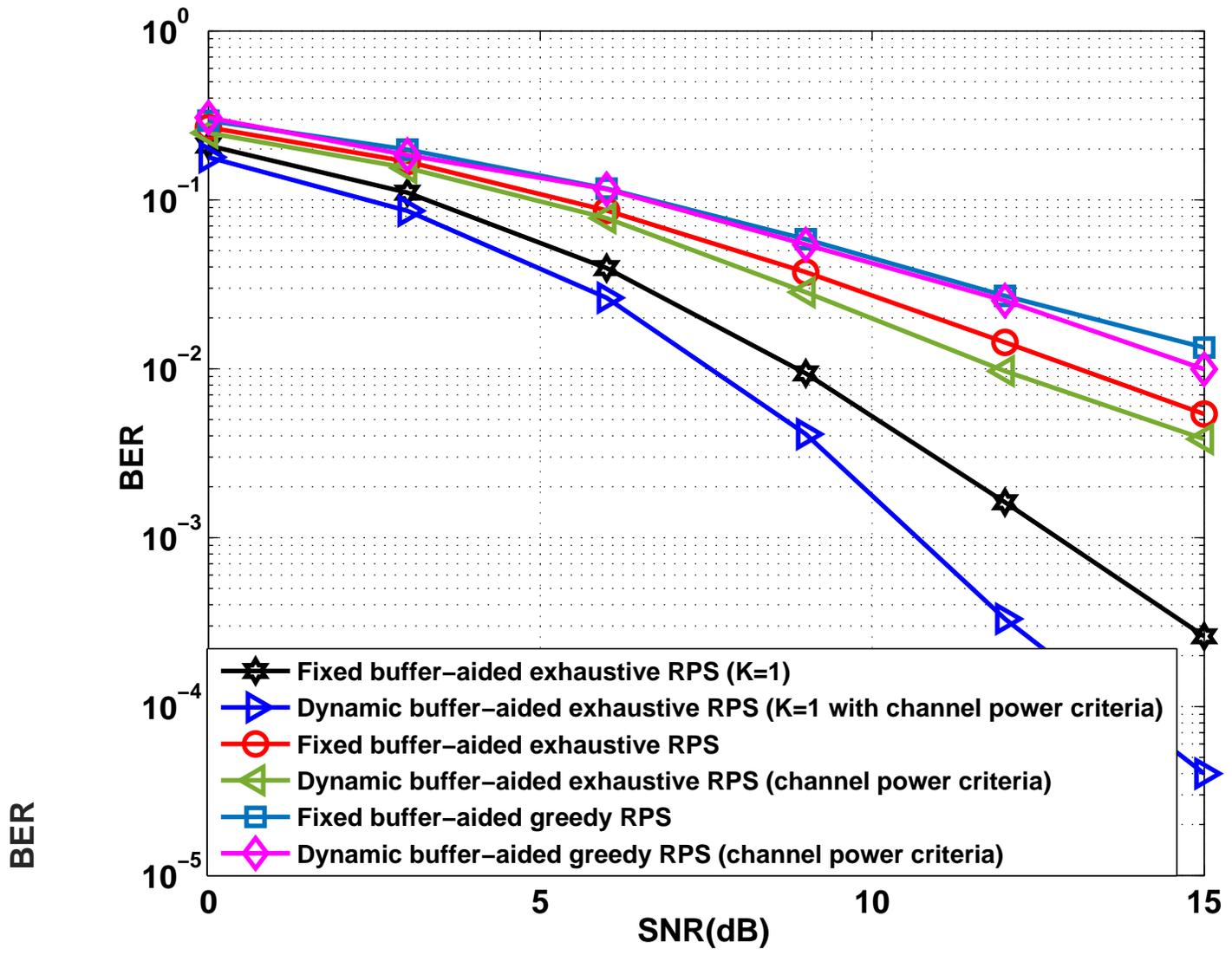
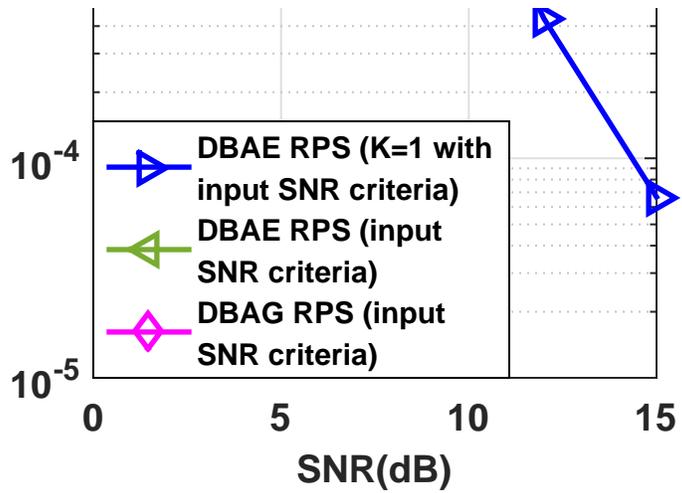
\end{document}